\documentclass[lettersize,journal]{IEEEtran}
\usepackage{amsmath,amsfonts}
\usepackage{algorithmic}
\usepackage{algorithm}
\usepackage{array}
\usepackage{booktabs}
\usepackage{makecell}
\usepackage{amsmath}
\usepackage{multirow}
\usepackage[caption=false,font=normalsize,labelfont=sf,textfont=sf]{subfig}
\usepackage{textcomp}
\usepackage{stfloats}
\usepackage{url}
\usepackage{verbatim}
\usepackage{graphicx}
\usepackage{cite}
\begin{document}

\title{LLM-Enhanced Deep Reinforcement Learning with Human-in-the-Loop Calibration for Real-Time Optimal Power Flow}

\author{Jiayang Tu, Bozhen Jiang, Bojun Du,~\IEEEmembership{Student Member,~IEEE,} Yunxiao Ling,\\Ershun Du,~\IEEEmembership{Member,~IEEE,} and Qin Wang,~\IEEEmembership{Senior Member,~IEEE}
\thanks{Jiayang Tu, Bozhen Jiang, Yunxiao Ling and Qin Wang are with the Department of Electrical and Electronic Engineering, The Hong Kong Polytechnic University, Hong Kong.}
\thanks{Ershun Du is with the Laboratory of Low Carbon Energy, Tsinghua University, Beijing 100084, China.}
\thanks{Bojun Du is with the Department of Electrical Engineering, Tsinghua University, Beijing 100084, China}
\thanks{Corresponding authors: Ershun Du and Qin Wang}
}

\maketitle

\begin{abstract}
Real-time optimal power flow (RT-OPF) requires fast and reliable decision-making under time-varying operating conditions. However, many practical operating preferences and safety related requirements are difficult to encode using conventional numerical rewards alone, which limits the effectiveness and controllability of deep reinforcement learning (DRL) for RT-OPF. To address this issue, this paper proposes an LLM-enhanced DRL framework with human-in-the-loop (HITL) for RT-OPF. The proposed method augments the physics-based reward with sparse language evaluative feedback during offline training, so that semantic operational requirements can be incorporated into policy learning. To improve reliability in RT-OPF, a selective HITL calibration mechanism is further introduced to inspect and correct questionable LLM judgments before they continuously shape the learning process. Case studies on the IEEE 30-bus and IEEE 118-bus systems show that LLM feedback improves policy safety over vanilla TD3, while further HITL calibration achieves the best overall performance and more stable training behavior. These results demonstrate the effectiveness of combining language guidance and expert calibration for RT-OPF.
\end{abstract}


\section{Introduction}
\IEEEPARstart{O}{ptimal} power flow (OPF) is a fundamental tool for power system operation, as it determines operating set-points that balance economy and security while satisfying network physics and engineering constraints \cite{Sayed2022}. As a large-scale nonconvex optimization problem, its computational complexity has long posed a major challenge to practical implementation \cite{Li2024}. In actual operation, however, OPF often needs to be solved repeatedly over short dispatch intervals in response to time-varying demand, operating condition changes, and intertemporal constraints \cite{Yi2023}. This has made real-time OPF (RT-OPF) increasingly important, while also imposing stringent requirements on computational efficiency and solution reliability for online decision-making \cite{Wu2024}.

To alleviate the burden of repeated online OPF solution, recent studies have explored learning assisted acceleration of model-based optimization as well as data-driven surrogates trained offline to approximate OPF solutions \cite{Rajaei2025,Lei2020}. However, model-driven acceleration still relies on iterative solver execution and accurate system models, while learned surrogates often struggle with feasibility and robustness across operating conditions \cite{Han2024}. In contrast, deep reinforcement learning (DRL) is well suited to RT-OPF because it provides a natural framework for learning control policies under uncertainty in continuous action spaces with temporal coupling \cite{Wu2023}. In contrast, DRL is particularly attractive for RT-OPF because it can learn control policies under uncertainty in continuous action spaces with temporal coupling, while enabling fast online inference after offline training \cite{Yan2022}. This is well aligned with OPF, where generator dispatch and voltage regulation are continuous and current actions affect future operating conditions and security margins. As a result, DRL and its safe variants have been increasingly applied to RT-OPF and security-constrained OPF \cite{Yan2020}. However, the effectiveness of such methods remains highly sensitive to environment construction and, in particular, to how reward signals encode operational objectives and requirements \cite{Wolgast2024}.

Although DRL is effective at optimizing policies from feedback, it does not inherently interpret semantic, preference-driven, or experience-based operating rules. This limitation is particularly relevant to OPF, where many practical operational requirements are not naturally expressible as standard numerical rewards or constraints, making conventional feedback engineering difficult \cite{Ma2024,Yan2023}. Therefore, an additional mechanism is needed to translate language-based requirements into structured training signals. This creates a natural role for large language models (LLMs), whose strengths in language understanding and reasoning make them suitable for interpreting high-level operational requirements \cite{Yasunaga2024}. Accordingly, LLM-enhanced RL literature \cite{Cao2024} has identified information processing and reward design as important interfaces through which LLMs can intervene in the RL loop. Such a gap is already visible in power system applications, where LLMs have been used to refine penalty functions for safe RL and inject forecast information and semantic grid knowledge into voltage control frameworks \cite{Yang2025,Yang2026}. Beyond grid control, recent literature \cite{Ma2025} suggest that LLM guidance can support sequential decision problems involving context-dependent preferences and high-level criteria. However, prior studies suggest that LLM-generated guidance alone may still be insufficient in safety-critical settings, because alignment with human intent remains difficult to ensure and model outputs may be unstable or unreliable \cite{yang2023large,Huang2023}.

In DRL training, the concern with LLM feedback is not only that it may hallucinate, but that once injected into the learning loop, it can reshape the signal from which the policy is optimized. Corrupted preference signals may systematically steer the learned policy away from its intended behavior \cite{Wang2024}. This tissue is further amplified in power system with safety as top priority. In addition, DRL are often regarded as black-box models whose internal policy evolution is difficult to inspect or verify directly \cite{Yu2025}. As a result, even when domain experts recognize inappropriate learned policies, conventional RL pipelines provide limited means to intervene in and correct the training process \cite{Kaufmann2025}. Prior work on safe and interactive RL has shown that explicit human intervention can improve safety and alignment during training, but such intervention is typically external to the reward construction and may remain costly or task-specific \cite{Xu2022,Peng2023}. In energy applications, expert involvement has traditionally relied on manual reward redesign, heuristic modification, or repeated hyperparameter adjustment \cite{Butt2024}, which are coarse-grained and labor-intensive, especially when the relevant requirements are semantic. From this perspective, LLMs can be viewed not only as generators of training feedback, but also as a linguistic interface through which domain experts can inspect and revise DRL training process. Related work further indicates that rule augmentation, demonstrations, and multi-turn interaction can make model-based assessment more controllable \cite{Li2025}. This view is also consistent with recent discussions in the power-system community that large foundation models can serve as a flexible interface between human knowledge and grid operation tasks \cite{Huang2024}.  Therefore, what is needed is a training mechanism in which language-based feedback remains reviewable, calibratable, and corrigible by domain experts before it persistently shapes policy learning.

To address the above issues, this paper proposes an LLM-enhanced deep reinforcement learning framework with human-in-the-loop (HITL) for real-time optimal power flow. In the proposed framework, LLMs are introduced during offline training to translate language-based operational requirements into auxiliary evaluative feedback, while domain experts are allowed to review and revise such feedback when necessary. The resulting calibrated signals are used to complement, rather than replace, the underlying physics-based reward design, so that the learned policy remains grounded in system operation objectives and constraints. After training, the final DRL policy can still be deployed online through direct forward inference, thereby preserving the computational advantage required by real-time OPF. The main contributions of this work are summarized as follows:

\textit{1) LLM-enhanced DRL with HITL framework:} We develop an LLM-enhanced DRL training framework with HITL for real-time OPF, in which language-based operational feedback is introduced during offline training. The trained policy remains directly deployable for fast online inference, thereby preserving the computational advantage required by real-time operation.

\textit{2) Language feedback calibration mechanism:} We develop an expert-calibration mechanism that enables domain experts to inspect, revise, and correct LLM-generated feedback during training, thereby improving the alignment between language-based guidance and actual operational intent.

\textit{3) Interpretable and corrigible training paradigm:} We establish an interpretable and corrigible training paradigm for OPF based DRL, transforming the learning process from a largely black-box optimization into a reviewable and adjustable loop without increasing online inference complexity.

\section{Problem Formulation}
\subsection{Real-Time OPF With Physical Constraints}
The RT-OPF problem determines a set of control decisions for each dispatch interval $t$. Let $\mathcal{N}$, $\mathcal{G}$, $\mathcal{L}$, and $\mathcal{T}$ denote the sets of buses, generators, transmission lines, and transformers, respectively. At each interval $t$, RT-OPF minimizes the total generation cost subject to the AC power flow equations and operational constraints by optimizing the controllable devices:
\begin{equation}
    \min\sum_{i\in\mathcal{G}}C_i(P_{G_{i}}^t)
\label{eq:1}
\end{equation}
where $P_{G_{i}}^t$ is the active power output of the $i$-th generator for the single time interval $t$; $C_i(\cdot)$ is the convex generation cost function in the IEEE bus system, which is modeled as a quadratic polynomial $C_i(P_{G_{i}}^t)=a_{i}(P_{G_{i}}^t)^2+b_{i}P_{G_{i}}^t+c_{i}$.

The operating constraints are power flow equations, generator active/reactive power output limits, bus voltage limits, transmission line power flow limits, and unit ramping limits:
\begin{equation}
    P_{G_{i}}^t-P_{D_{i}}^t=V_i^t\sum_{j\in\mathcal{N}}V_j^t\left(G_{ij}\cos\theta_{ij}^t + B_{ij}\sin\theta_{ij}^t\right)
\label{eq:2}
\end{equation}
\begin{equation}
    Q_{G_{i}}^t-Q_{D_{i}}^t=V_i^t\sum_{j\in\mathcal{N}}V_j^t\left(G_{ij}\sin\theta_{ij}^t - B_{ij}\cos\theta_{ij}^t\right)
\label{eq:3}
\end{equation}
\begin{equation}
    P_{G_{i}}^{\min} \le P_{Gv}^t \le P_{G_{i}}^{\max}, \forall i\in\mathcal{G}
\label{eq:4}
\end{equation}
\begin{equation}
    Q_{G_{i}}^{\min} \le Q_{G_{i}}^t \le Q_{G_{i}}^{\max}, \forall i\in\mathcal{G}
\label{eq:5}
\end{equation}
\begin{equation}
    V_i^{\min} \le V_i^t \le V_i^{\max},\forall i\in\mathcal{N}
\label{eq:6}
\end{equation}
\begin{equation}
    |S_{\ell}^t| \le S_{\ell}^{\max},\forall l\in\mathcal{L}
\label{eq:7}
\end{equation}
\begin{equation}
    |P_{G_{i}}^t - P_{G_{i}}^{t-1}| \le R_i\Delta t,\forall i\in\mathcal{G}.
\label{eq:8}
\end{equation}
where $P_{G_{i}}^t$ and $Q_{G_{i}}^t$ are the active and reactive power outputs of the generator at bus $i$ at time $t$; 
$P_{D_{i}}^t$ and $Q_{D_{i}}^t$ are the active and reactive power demands at bus $i$ at time $t$; 
$V_i^t$ and $V_j^t$ are the voltage magnitude at bus $i$ and $j$ at time $t$;
$\theta_{ij}^t$ is the voltage angle difference between buses $i$ and $j$ at time $t$; 
$G_{ij}$ and $B_{ij}$ denote the conductance and susceptance terms of the $(i,j)$-th entry of the bus admittance matrix $Y_{ij}=G_{ij}+jB_{ij}$, respectively; 
$P_{G_{i}}^{\min}$ and $P_{G_{i}}^{\max}$ are the lower and upper limits of generator active power output;
$Q_{G_{i}}^{\min}$ and $Q_{G_{i}}^{\max}$ are the lower and upper limits of generator reactive power output; $V_i^{\min}$ and $V_i^{\max}$ are the lower and upper limits of the bus voltage magnitude;
$S_l^t$ is the apparent power flow on line $l$ at time $t$ and $S_l^{\max}$ is its thermal limit;
$R_i$ is the ramp-rate limit of generator $i$ (in power per unit time);
and $\Delta t$ is the length of a dispatch interval.

The above formulation characterizes the economic objective and the physical operating constraints that can be explicitly specified in analytical form. In practical real-time operation, however, not all relevant operational requirements are represented in this way. Some requirements are conveyed in linguistic form and are not directly incorporated into the conventional RT-OPF formulation. 

\subsection{Language-based Operational Constraints}

Beyond the physical constraints in \eqref{eq:2}--\eqref{eq:8}, practical real-time operation may also involve requirements that are specified in linguistic rather than analytical form. These requirements do not replace the conventional RT-OPF objective of minimizing operating cost, but supplement it by expressing how candidate operating points should be assessed from an operational perspective. In this sense, they act as language-based operational constraints on the learning process: they indicate what kinds of behaviors are preferred, discouraged, or regarded as operationally inappropriate, even when such requirements are not directly prescribed as fixed mathematical expressions.

\begin{figure}[!t]
\centering
\includegraphics[width=\columnwidth]{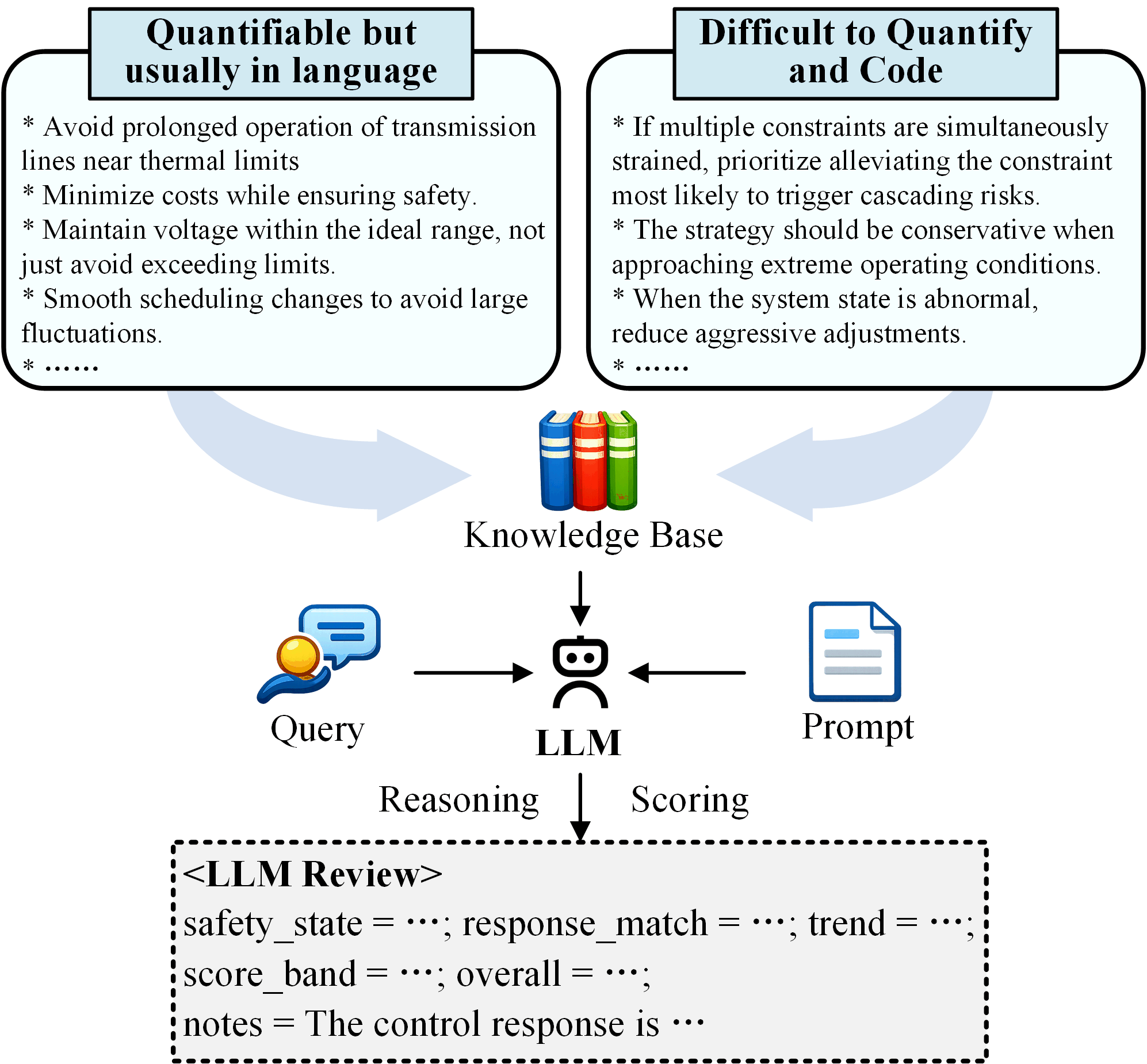}
\caption{Translation of natural-language rules into structured evaluation.}
\label{fig:Rules to Reward}
\vspace{-0.25cm}
\end{figure}

As illustrated in Fig.~\ref{fig:Rules to Reward}, the language-based operational constraints considered in this work can be grouped into two categories. The first category consists of quantifiable rules that are usually expressed in language. These requirements correspond to identifiable operating attributes, but are often described qualitatively in practice rather than prescribed as fixed numerical criteria. The second category consists of rules that are difficult to quantify and code. Although their operational intent may be clear to human operators, their direct translation into conventional mathematical constraints or reward terms is difficult because the evaluation depends on context and may involve interactions among multiple constraints. Therefore, while these requirements are relevant to RT-OPF decision making, they are not directly usable by optimization or learning algorithms in their original natural language form.

To make such requirements usable in training, an intermediate interpretation mechanism is needed to transform them into structured signals that can interact with the learning objective. In this work, LLM serves this role. Specifically, the LLM takes as input the language-based operational requirements together with the current operating outcome, and converts them into structured evaluative feedback. The converted feedback is not intended to substitute the physics constraints; instead, it provides an auxiliary assessment layer through which language-based requirements can influence policy learning. In this way, the semantic content of operational requirements can be incorporated into the training process for solving the RT-OPF problem, while the underlying optimization target remains grounded in economic performance and physical feasibility.

\section{Proposed Method}
This section presents the proposed LLM-enhanced DRL framework with HITL calibration for RT-OPF. Based on the formulation in Section II, the proposed method integrates a physics-based DRL backbone, an LLM evaluative agent, and a selective expert calibration mechanism. The overall framework is first introduced, followed by the DRL formulation, the LLM agent, and the HITL calibration process.

\subsection{Overall Framework of the Proposed Method}

\begin{figure*}[!t]
\centering
\includegraphics[width=\textwidth]{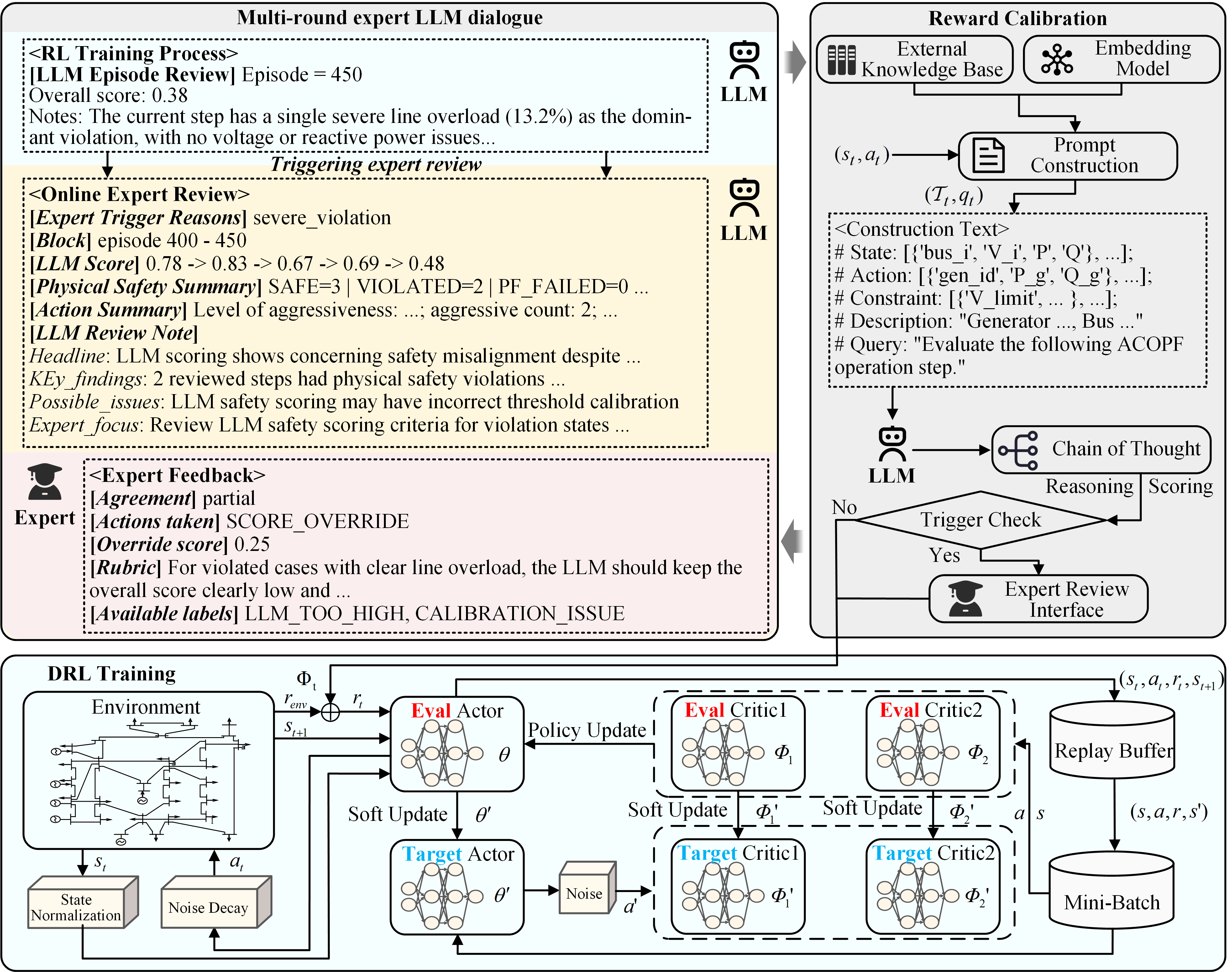}
\caption{Overall framework of the proposed LLM-enhanced DRL with human-in-the-loop for RT-OPF.}
\label{fig:framework}
\end{figure*}

The proposed method adopts the LLM-enhanced DRL framework shown in Fig.~\ref{fig:framework} for RT-OPF, where the physical power system simulator, a DRL control backbone, and an LLM-based semantic evaluation module are integrated into a unified training loop. Specifically, the RT-OPF environment provides state transitions and physics-based reward according to the network operating outcome, while the DRL agent learns a continuous control policy for generator dispatch. In parallel, the LLM agent evaluates the realized operating behavior against language-based operational constraints and produces an auxiliary shaping signal, which can be selectively calibrated by a human expert when necessary.

\begin{algorithm}[t]
\caption{LLM-Enhanced DRL Training Pipeline}
\label{alg:overall_framework}
\begin{algorithmic}[1]
\STATE \textbf{Initialize:} RT-OPF environment, TD3 agent, replay buffer $\mathcal{D}$, LLM agent $\mathcal{M}$, expert calibration module $\mathcal{H}$, scoring interval $K$, shaping signal cache $\Phi \leftarrow 0$
\FOR{$t=1$ to $T$}
    \STATE \textit{// DRL-environment interaction}
    \STATE Observe state $s_t$
    \STATE Select continuous OPF action $a_t$ with exploration noise
    \STATE Execute $a_t$ in the RT-OPF environment and observe envrionment reward $r_t^{\mathrm{env}}$, next state $s_{t+1}$, and done flag $d_t$

    \STATE \textit{// LLM-based scoring}
    \STATE Initialize shaping signal $\Phi_t$ from the recent LLM score
    \IF{$t \bmod K = 0$}
        \STATE Query the LLM agent $\mathcal{M}$ with the current state, action, and physical operating outcome
        \STATE Obtain LLM score $\Phi_t^{\mathrm{LLM}}$

        \STATE \textit{// Human expert calibration}
        \IF{expert review is triggered}
            \STATE Request expert feedback from $\mathcal{H}$
            \STATE Calibrate LLM score or scoring rubric
            \STATE Obtain calibrated shaping signal $\Phi_t$
        \ELSE
            \STATE Set $\Phi_t \leftarrow \Phi_t^{\mathrm{LLM}}$
        \ENDIF
    \ENDIF

    \STATE \textit{// Reward integration and TD3 update}
    \STATE Form final training reward:
    \[
    r_t \leftarrow \beta r_t^{\mathrm{env}} + \alpha \Phi_t
    \]
    \STATE Store transition $(s_t,a_t,r_t,s_{t+1},1-d_t)$ in $\mathcal{D}$
    \STATE Sample a mini-batch from $\mathcal{D}$ and update the TD3 critics
    \IF{$t \bmod d = 0$}
        \STATE Update the TD3 actor and target networks
    \ENDIF
\ENDFOR
\end{algorithmic}
\end{algorithm}

The overall training pipeline is summarized in Algorithm~\ref{alg:overall_framework}. In this work, TD3 \cite{fujimoto2018} is adopted as the DRL backbone due to its suitability for continuous-control problems and its stable actor--critic training mechanism. At each interaction step, the agent observes the current state, selects a continuous control action with exploration noise, and executes it in the environment to obtain the environment reward, next state, and termination flag. Meanwhile, the framework maintains a cache from the most recent valid LLM score. Every $K$ training steps, the current state, action, and operating outcome are submitted to the LLM agent for an updated semantic score. If expert review is triggered, human feedback is used to calibrate the score or revise the scoring rubric before the shaping signal is fused with the environment reward to form the final training reward. The resulting transition is then stored in the replay buffer, and mini-batches are sampled to update the TD3 backbone.

\subsection{DRL Formulation for RT-OPF}
The RT-OPF control task is formulated as a Markov decision process (MDP), so that a closed-loop policy can be learned to map time-varying system conditions to real-time dispatch decisions. At each dispatch interval $t$, the agent observes the current operating state $s_t$, selects a continuous control action $a_t$, and applies it to the RT-OPF environment. The power system then evolves to the next operating point under network physics and exogenous load variations, resulting in the next state $s_{t+1}$. The MDP components are defined as follows.

\textit{1) State Space $\mathcal{S}$:} The system state at time $t$, denoted as $s_t\in\mathcal{S}$, summarizes the real-time operating condition of the power network for RT-OPF dispatch. In this paper, $s_t$ is constructed from the active/reactive load demands at time $t$ and the generator active power dispatch from the previous interval:
\begin{equation}
s_t = \{P_D^t,\ Q_D^t,\ P_G^{t-1}\}
\label{eq:state_space}
\end{equation}

\textit{2) Action Space $\mathcal{A}$:} The system action at time $t$, denoted as $a_t\in\mathcal{A}$, specifies the real-time setpoints of controllable devices for RT-OPF dispatch. In this paper, the action is defined as a continuous vector that concatenates the generator active power setpoints and the generator voltage magnitude setpoints:
\begin{equation}
a_t = \{P_G^t,\ V_G^t\}
\label{eq:action_space}
\end{equation}

\textit{3) Transition Probability $\mathcal{P}(s_{t+1}\mid s_t,a_t)$:} The state transition function $\mathcal{P}:\mathcal{S}\times\mathcal{A}\times\mathcal{S}\rightarrow [0,1]$ specifies the probability of transitioning to the next state $s_{t+1}$ from a given state $s_t$ under action $a_t$. In RT-OPF, the transition dynamics are primarily governed by the nonlinear AC power flow equations together with exogenous load variations. Since this transition model is not available in closed form, a model-free formulation is adopted, where $\mathcal{P}$ is implicit in the environment simulator.

\textit{4) Environment Reward:} The physics-based environment reward is designed to align the learning objective with the RT-OPF operational goal, i.e., minimizing generation cost while discouraging violations of critical operating constraints. At each interval $t$, the quadratic generation cost $C_t$ is computed, together with a set of nonnegative violation metrics $\{v_k^t\}_{k\in\mathcal{K}}$ associated with line overloads, bus-voltage limit violations, generator reactive power limit violations, and ramping violations. A primal-dual Lagrangian shaping is then applied by introducing nonnegative multipliers $\{\lambda_k^t\}_{k\in\mathcal{K}}$, yielding in the shaped stage cost and corresponding environment reward:
\begin{equation}
r_t^{\mathrm{env}}=-J_t = -\left(W_\mathrm{gen}\,C_t + \sum_{k\in\mathcal{K}} W_k\,\lambda_k^t\, v_k^t\right)
\label{eq:environment_reward}
\end{equation}
where $\mathcal{K}=\{\mathrm{line},\mathrm{volt},q,\mathrm{ramp}\}$ and $W_{\mathrm{gen}},W_k$ are scalar weights.

The above formulation defines the physics-based learning backbone of the proposed method. For a sampled transition batch $\{(s_i,a_i,r_i,s_i',d_i)\}_{i=1}^{N}$, where $s_i$, $a_i$, $r_i$, $s_i'$, and $d_i$ denote the state, action, reward, next state, and terminal indicator of the $i$-th sampled transition, respectively, the smoothed target action is constructed as
\begin{equation}
\tilde a_i = \pi_{\theta'}(s_i')+\epsilon,\qquad
\epsilon\sim \mathrm{clip}(\mathcal N(0,\tilde\sigma),-c,c),
\label{eq:td3_target_action}
\end{equation}
where $\pi_{\theta'}$ denotes the target actor network parameterized by $\theta'$; $\epsilon$ is the clipped Gaussian perturbation used for target policy smoothing, with standard deviation $\tilde{\sigma}$ and clipping threshold $c$. The critic target is then computed by
\begin{equation}
y_i = r_i + \gamma(1-d_i)\min_{j=1,2}Q_{\phi_j'}(s_i',\tilde a_i)
\label{eq:td3_target}
\end{equation}
where $\gamma\in(0,1)$ is the discount factor; $Q_{\phi_1'}$ and $Q_{\phi_2'}$ are the target critic networks parameterized by $\phi_1'$ and $\phi_2'$, respectively, and the minimum operator is used to mitigate overestimation bias. Based on the target value, the twin critics are updated by minimizing
\begin{equation}
\mathcal L_Q
=
\frac{1}{N}\sum_{i=1}^{N}\sum_{j=1}^{2}
\left(y_i-Q_{\phi_j}(s_i,a_i)\right)^2
\label{eq:td3_critic_loss}
\end{equation}
where $\mathcal L_Q$ denotes the critic loss. The actor is updated using the deterministic policy gradient
\begin{equation}
\nabla_\theta J(\theta)
=
\frac{1}{N}\sum_{i=1}^{N}
\nabla_a Q_{\phi_1}(s_i,a)\big|_{a=\pi_\theta(s_i)}
\nabla_\theta \pi_\theta(s_i),
\label{eq:td3_actor_update}
\end{equation}
which encourages the current policy to move toward actions with higher critic estimated return. To stabilize training, the target networks are softly updated according to
\begin{equation}
\phi_j' \leftarrow \tau \phi_j + (1-\tau)\phi_j',\qquad j\in\{1,2\},
\label{eq:td3_target_critic_update}
\end{equation}
\begin{equation}
\theta' \leftarrow \tau \theta + (1-\tau)\theta'.
\label{eq:td3_target_actor_update}
\end{equation}
where $\tau\in(0,1)$ is the soft update rate.

It should be emphasized that the above TD3 updates only constitute the physics-based continuous-control backbone of the proposed method. The language-based operational constraints introduced in Section II-B are not directly imposed as additional hard mathematical constraints in the MDP. Instead, they are incorporated through an auxiliary LLM-based evaluative mechanism, which is described in the next subsection.

\subsection{LLM Agent Design}
The LLM module in this work is not introduced as an RT-OPF solver, nor is it used to replace physical power flow computation. Instead, it serves as an intermediate semantic layer that interprets language-based operational constraints and converts them into structured evaluative feedback for DRL training. Given the language-based operational constraints introduced in Section II-B, the role of the LLM is to assess the realized operating behavior with respect to such requirements and provide an auxiliary semantic signal that complements the physics-based environment reward.

\begin{table}[!t]
\caption{Two Types of Prompt Design Categories}
\label{tab:prompt design}
\centering
\footnotesize
\renewcommand{\arraystretch}{1.5}
\setlength{\tabcolsep}{4pt}
\begin{tabular}{p{0.11\columnwidth} p{0.40\columnwidth} p{0.39\columnwidth}}
\toprule
\textbf{Category} & \textbf{Evaluation Prompt} & \textbf{Expert Review Prompt} \\
\midrule
Role
& You are OPFSafetyJudger, an expert evaluator for one current real-time ACOPF control step with recent training context.
& You are an expert-review assistant for an ACOPF deep reinforcement learning training pipeline. \\

Task
& Evaluate the quality of ONE CURRENT ACOPF control step using ...
& Your role is to review a recent block summary of ... \\

Context
& Use only the provided JSON context for this evaluation ...
& Use only the provided block summary of ... \\

Rules
& Your judgment must be safety firstly ...
& Use only the provided block summary ... \\

Format
& Return ONLY one valid JSON object. The JSON object MUST contain ...
& Return ONLY valid JSON with this schema ... \\

Example
& \makecell[tl]{\# Example\\Current step: ...\\Expected judgment: ...}
& None \\

CoT
& Decision Procedure: (1) Determine safety state ...; (2) Determine the dominant violation ...; (3) Determine response matches ...; (4) Determine the short-horizon trend ...; (5) Choose a score band ...; (6) Determine the final score ...
& The expert note should help the expert quickly identify: (1) Judge overall scoring reasonableness ...; (2) Judge scoring tendency ...; (3) Determine Consistency ...; (4) Determine actions patterns ...; (5) Determine trend window ...; (6) Give inspection priority ... \\

\bottomrule
\end{tabular}
\end{table}

To operate this semantic evaluation process, Table~\ref{tab:prompt design} defines constrained LLM prompts for structured operational evaluation. Its purpose is not to elicit unconstrained narrative responses from the LLM, but to stabilize the judgment scope, reasoning path, and output schema for downstream training use. The main design considerations are as follows.

\textit{1) Role:} The role field anchors the LLM to a task-specific identity, thereby reducing behavioral drift and keeping the model aligned with the intended evaluation objective.

\textit{2) Task:} The task field specifies the temporal and functional scope of each prompt, distinguishing current-step evaluation from block review inspection.

\textit{3) Context:} The context field restricts the admissible evidence source, which reduces unsupported extrapolation and improves the traceability of the LLM judgment.

\textit{4) Rules:} The rules field injects domain-specific principles that should not be left unconstrained, especially the safety-first criterion in step-level evaluation and the grounded diagnostic requirement in expert review.

\textit{5) Format:} Both prompts require exactly one valid JSON object as output, which converts the LLM response into a machine-readable interface for deterministic parsing in reward shaping, logging, trigger detection, and expert inspection.

\textit{6) Example:} Representative examples are mainly included in the evaluation prompt to illustrate the mapping from operating evidence to structured judgment output. This design leverages the few-shot learning capability of large language models~\cite{Brown2020} to stabilize scoring granularity and output style.

\textit{7) CoT:} The CoT component is implemented as an ordered reasoning scaffold rather than unrestricted free-form thinking. This follows the general idea of chain-of-thought prompting, which improves reasoning quality by decomposing complex tasks into intermediate steps~\cite{Wei2022}.

In this way, the LLM module serves as a stable semantic interface connecting language-based operational constraints, structured evaluation, DRL reward shaping, and subsequent HITL calibration. Based on the structured evaluation returned by the LLM, a semantic score is generated for the current operating step. Let $\phi_t^{\mathrm{LLM}}$ denote the raw LLM score produced for the state--action--outcome tuple at step $t$. Since the LLM is queried sparsely rather than at every step, the score is updated only at designated scoring instants and is otherwise inherited from the most recent valid evaluation. Denoting the set of scoring instants by $\mathcal{T}_{\mathrm{score}}$, the raw LLM score is written as
\begin{equation}
\phi_t^{\mathrm{LLM}}=
\begin{cases}
g_{\mathrm{LLM}}(s_t,a_t,o_t), & t\in\mathcal{T}_{\mathrm{score}}\\
\phi_{t^-}^{\mathrm{LLM}}, & t\notin\mathcal{T}_{\mathrm{score}}
\end{cases}
\label{eq:llm_raw_score}
\end{equation}
where $g_{\mathrm{LLM}}(\cdot)$ denotes the structured step-level evaluation procedure defined by the prompt interface in this subsection, $o_t$ denotes the realized physical operating outcome after executing action $a_t$ under state $s_t$, and $\phi_{t^-}^{\mathrm{LLM}}$ is the most recent valid LLM score.

\subsection{Reward Shaping and Human Expert Calibration}
Although the LLM provides a structured semantic evaluation channel for language-based operational constraints, its judgment is not guaranteed to be fully reliable in safety-critical scenarios. In RT-OPF, once an inaccurate semantic score is injected into the reward pathway, it may continuously influence policy updates and thereby distort the learning direction. Therefore, the proposed method introduces a HITL calibration mechanism to selectively inspect and correct questionable LLM feedback before it is used for policy learning. The role of HITL is not to intervene at every training step. Instead, expert involvement is activated only when the reliability of the LLM judgment becomes questionable. As illustrated in Fig.~\ref{fig:framework}, such cases typically include severe or repeated physical violations, persistent low-score patterns, or mismatch between the LLM semantic score and the realized physical operating outcome. Under these conditions, a recent review block is constructed from the training history to summarize recent state--action--outcome tuples, LLM scores, physical safety outcomes, and their consistency. This block is then provided to the expert review to support efficient human inspection.

To distinguish the raw LLM score from the effective shaping term finally injected into training, let $\Phi_t$ denote the calibrated semantic shaping signal used in the reward. In the absence of expert intervention, the effective shaping signal is directly determined by the raw LLM score. When HITL is triggered, the expert feedback modifies the shaping term through a calibration operation. Accordingly, $\Phi_t$ is defined as
\begin{equation}
\Phi_t=
\begin{cases}
\phi_t^{\mathrm{LLM}}, & \text{if no HITL is triggered}\\
\mathcal{C}\!\left(\phi_t^{\mathrm{LLM}},\mathcal{E}_b\right), & \text{if HITL is triggered}
\end{cases}
\label{eq:effective_phi}
\end{equation}
where $\mathcal{E}_b$ denotes the expert feedback associated with review block $b$, and $\mathcal{C}(\cdot)$ denotes the corresponding calibration operation. In practice, this calibration may either directly adjust the current semantic score or refine the scoring rubric used for subsequent LLM evaluations.

Based on the calibrated shaping signal, the final training reward is formed as
\begin{equation}
r_t = \beta\, r_t^{\mathrm{env}} + \alpha\, \Phi_t
\label{eq:final_reward_general}
\end{equation}
where $\beta$ and $\alpha$ are weighting coefficients, and $\Phi_t$ denotes the effective semantic shaping term injected into the training reward. Therefore, the LLM-derived semantic feedback should be understood as an auxiliary training assessment signal rather than a replacement for the underlying physical objective and hard operating constraints.

\section{Case Study}
\subsection{Experimental Setup}
\label{subsec:exp_setup}
The proposed framework is evaluated on the IEEE 30-bus and IEEE 118-bus transmission systems implemented in \texttt{pandapower} \cite{thurner2018}. The IEEE 30-bus case is mainly used for mechanism analysis and representative physical-behavior visualization, whereas the IEEE 118-bus case is further employed to verify the effectiveness of the proposed framework on a larger transmission system. The DRL framework is implemented in Python 3.8.20 using PyTorch 2.4.1 with cu121.

The time-varying operating scenarios are constructed from hourly load profiles derived from AC power flow data generated following the PowerModels.jl framework \cite{coffrin2018}. These hourly load sequences are further rescaled such that the resulting total demand level matches the nominal loading condition of the IEEE 30-bus and IEEE 118-bus systems. One day is defined as one episode, and each episode contains 24 dispatch steps corresponding to 24 hourly operating points. In this way, the sequential RT-OPF problem is cast as a daily decision-making process with temporally coupled operating states.

\begin{table}[h]
\caption{Hyperparameter Settings of the LLM Agent}
\label{tab:llm_hparams}
\centering
\footnotesize
\renewcommand{\arraystretch}{1.15}
\setlength{\tabcolsep}{10pt}
\begin{tabular}{ll}
\toprule
\textbf{Parameters} & \textbf{Value} \\
\midrule
Model                    & deepseek-v3.2 \\
Temperature              & 0.1 \\
Max tokens               & 1024 \\
\bottomrule
\end{tabular}
\end{table}

\begin{table}[h]
\caption{Hyperparameter Settings of the DRL Agent}
\label{tab:drl_hparams}
\centering
\footnotesize
\renewcommand{\arraystretch}{1.15}
\setlength{\tabcolsep}{6pt}
\begin{tabular}{p{0.40\columnwidth} p{0.35\columnwidth}}
\toprule
\textbf{Parameters} & \textbf{Value (30 bus/118 bus)} \\
\midrule
Optimizer                        & Adam \\
Activation function              & SiLU \\
Discount factor                  & 0.99 / 0.95 \\
Soft update rate                 & 0.005\\
Policy noise                     & 0.05 / 0.03 \\
Noise clip                       & 0.10 / 0.05 \\
Policy update frequency          & 2 \\
MLP hidden dimension             & 128 / 256\\
Number of hidden layers          & 3 \\
Actor learning rate              & 5e-5 / 1e-4\\
Critic learning rate             & 5e-5 / 5e-4 \\
Total training steps             & 5e4 \\
Exploration steps                & 2e3 / 4e3 \\
Replay buffer size               & 2e5\\
Batch size                       & 256\\
Initial exploration noise        & 1e-2\\
Final exploration noise          & 1e-3 \\
Learning rate decay              & Exponential\\
\bottomrule
\end{tabular}
\end{table}

Three learning configurations are considered in this study:

\textit{1) TD3:} the vanilla DRL backbone trained only with the physics-based environment reward;

\textit{2) with LLM:} the TD3 backbone augmented with language-based evaluative feedback;

\textit{3) with HITL:} the LLM-enhanced TD3 further equipped with selective human calibration.

The hyperparameter settings of the LLM agent and the DRL agent are summarized in Tables~\ref{tab:llm_hparams} and \ref{tab:drl_hparams}, respectively. Unless otherwise specified, all experiments use a fixed random seed. After offline training, the learned policies are fixed and evaluated on 10 test episodes.

The performance of different methods is evaluated using the operational cost, overall violation rate, and constraint average violation value. Let
$N^{\mathrm{line}}_{\mathrm{viol},t}$,
$N^{\mathrm{volt}}_{\mathrm{viol},t}$,
$N^{\mathrm{qgen}}_{\mathrm{viol},t}$, and
$N^{\mathrm{ramp}}_{\mathrm{viol},t}$
denote the numbers of violated line load rate, voltage of bus, generator reactive power, and generator ramp constraints at step $t$, respectively. The step overall violation rate is defined as
\begin{equation}
\mathrm{OVR}_t =
\frac{
N^{\mathrm{line}}_{\mathrm{viol},t}
+
N^{\mathrm{volt}}_{\mathrm{viol},t}
+
N^{\mathrm{qgen}}_{\mathrm{viol},t}
+
N^{\mathrm{ramp}}_{\mathrm{viol},t}
}{
|\mathcal{L}| + |\mathcal{N}| + |\mathcal{G}| + |\mathcal{G}|
},
\label{eq:ovr}
\end{equation}
where $|\mathcal{L}|$, $|\mathcal{N}|$, and $|\mathcal{G}|$ denote the numbers of transmission lines, buses, and generators, respectively. 

As an optimization reference, a conventional PF is obtained by solving the current network snapshot with the \texttt{pandapower} OPF solver at each dispatch step. This reference is used only for getting training states and is not involved in the policy learning process.

\subsection{Training Dynamics Under Different Feedback Mechanisms}
\label{subsec:training_dynamics}

\begin{figure}[!t]
\centering
\includegraphics[width=\columnwidth]{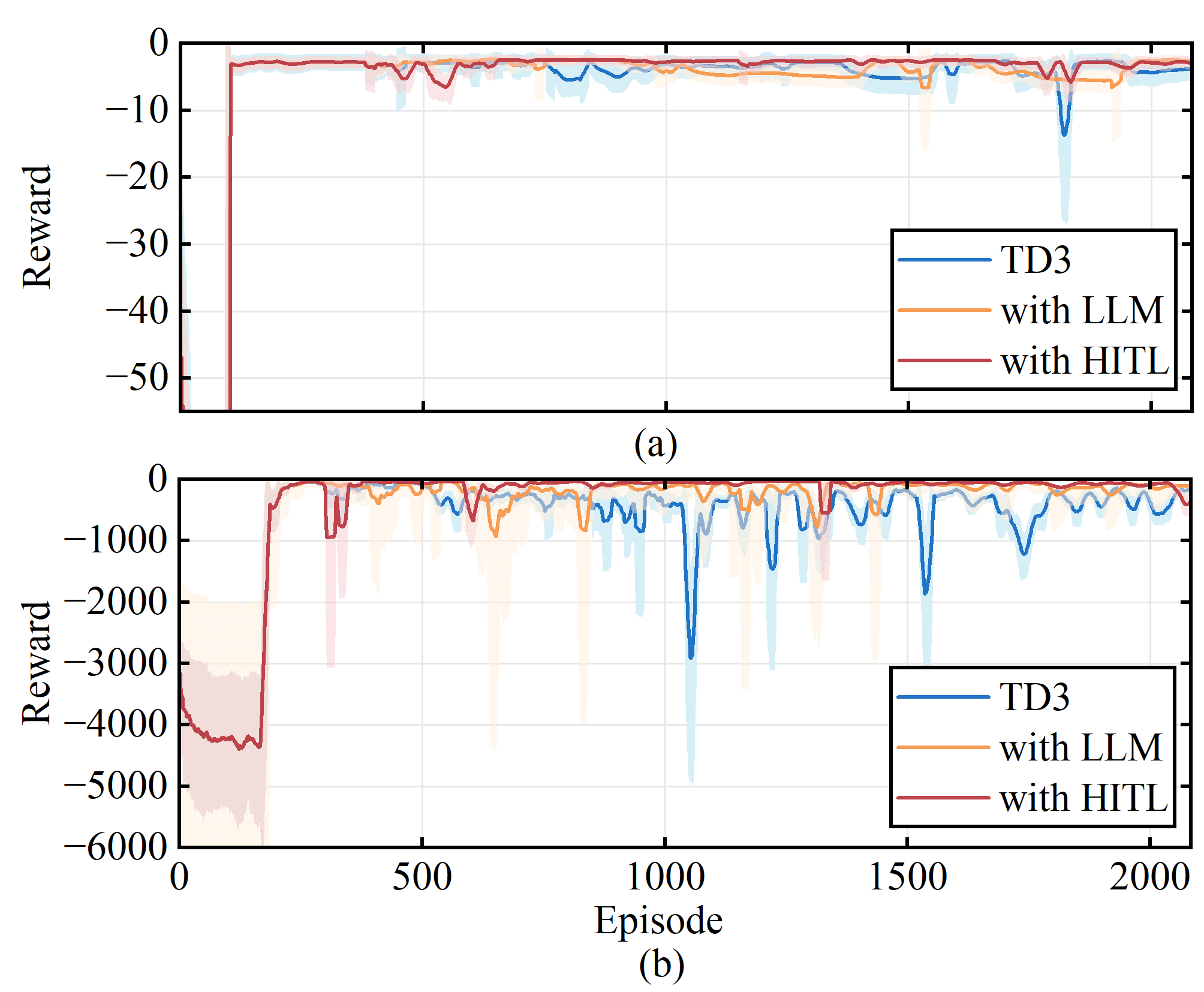}
\caption{Training reward trajectories under different feedback mechanisms: (a) 30-bus system; (b) 118-bus system.}
\label{fig:Reward}
\end{figure}

\begin{figure}[!t]
\centering
\includegraphics[width=\columnwidth]{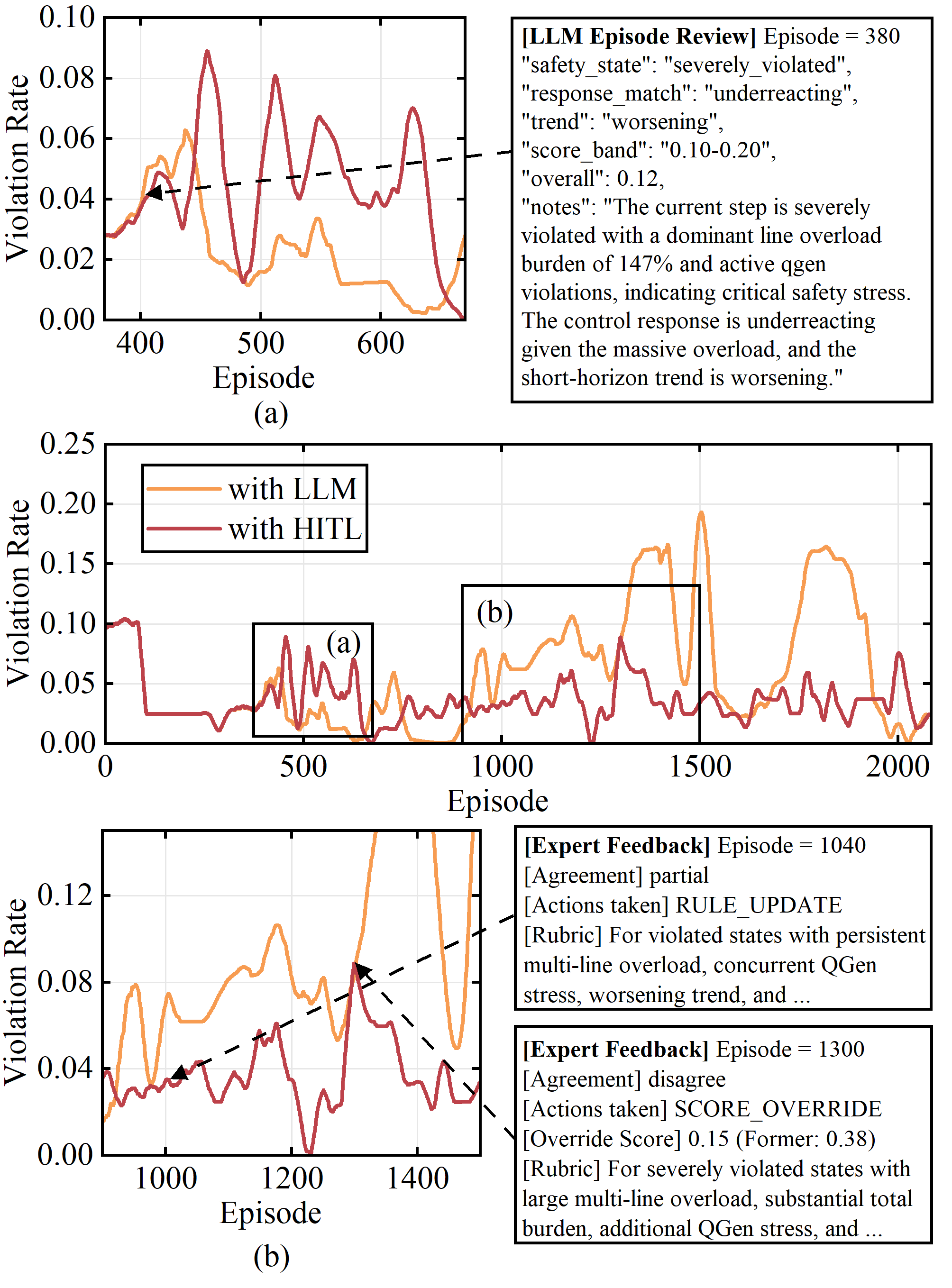}
\caption{Training violation trajectories on the 30-bus system for \textit{with LLM} and \textit{with HITL} with representative LLM review and HITL calibration examples.}
\label{fig:30trainvio}
\end{figure}

The training dynamics of the compared methods are illustrated from two complementary perspectives. Fig.~\ref{fig:Reward} presents the reward trajectories of \textit{TD3}, \textit{with LLM}, and \textit{with HITL}, so as to show how language-based feedback changes the overall learning process relative to the vanilla DRL backbone. By contrast, Fig.~\ref{fig:30trainvio} focuses specifically on the violation trajectories of \textit{with LLM} and \textit{with HITL}, together with representative LLM review and expert calibration examples, in order to highlight the additional role of HITL in correcting questionable LLM judgments during training.

As shown in Fig.~\ref{fig:Reward}, introducing language-based feedback changes the reward evolution relative to vanilla TD3, indicating that semantic evaluation meaningfully affects the policy optimization process. Compared with \textit{TD3}, the \textit{with LLM} setting exhibits a different training trajectory, suggesting that the language-based evaluative signal provides additional guidance beyond the physics-based reward. Moreover, \textit{with HITL} further improves the training behavior by making the reward evolution more stable in later stages, suggesting that expert calibration improves the stability of language-based policy learning in later training stages.

The effect of expert calibration is more clearly reflected in Fig.~\ref{fig:30trainvio}. In the early stage of training, the violation rate of \textit{with HITL} is not always lower than that of \textit{with LLM}, because the policy is still undergoing rapid exploration and adjustment, and the calibration mechanism may temporarily preserve or even amplify conservative corrections when the learned control responses have not yet stabilized. However, as training proceeds, the difference becomes increasingly evident. The \textit{with LLM} setting shows several periods of renewed violation increase after temporary improvement, indicating that the LLM semantic feedback may occasionally misjudge high-risk operating conditions and introduce misleading shaping signals due to hallucination. By contrast, \textit{with HITL} maintains a consistently lower and smoother violation trajectory in the later stage, showing that expert calibration effectively suppresses the accumulation of unreliable semantic judgments.

The representative examples in Fig.~\ref{fig:30trainvio} further illustrate this mechanism. Under severely violated operating conditions, the LLM may underestimate the severity of the current control response. If such underestimation is repeatedly injected into the reward, the learned policy may gradually deviate from the intended safety direction. The expert feedback shows that the calibration mechanism can either override the current score or revise the scoring rubric for subsequent evaluations. As a result, the semantic feedback channel becomes reviewable and corrigible, which is the key reason why \textit{with HITL} further improves the reliability of LLM-guided training.

\subsection{Test Performance and Physical Operation Characteristics}
\label{subsec:final_performance}

After offline training, the learned policies are fixed and evaluated on the test episodes. The overall test performance on the IEEE 30-bus and IEEE 118-bus systems is summarized in Tables~\ref{tab:performance_30bus} and \ref{tab:performance_118bus}, respectively. On both systems, introducing language feedback reduces the overall violation rate compared with the TD3, indicating that the semantic guidance is beneficial not only during training but also in the final learned policy. Moreover, the HITL calibration achieves the lowest overall violation rate among the learned methods on both systems, showing that expert calibration further improves the safety of the resulting policy.From the economic perspective, the \textit{with LLM} method attains the lowest operational cost among the learned policies, whereas the \textit{with HITL} method trades a limited amount of economic performance for more reliable safety improvement. At the same time, \textit{with HITL} consistently yields the smallest cost standard deviation, which indicates lower operating variability and a more stable policy output.

\begin{table}[!t]
\caption{Final test performance on 30-bus system}
\label{tab:performance_30bus}
\centering
\footnotesize
\renewcommand{\arraystretch}{1.2}
\setlength{\tabcolsep}{4pt}
\begin{tabular}{ 
>{\centering\arraybackslash}m{0.16\columnwidth} >{\centering\arraybackslash}m{0.165\columnwidth} >{\centering\arraybackslash}m{0.165\columnwidth} >{\centering\arraybackslash}m{0.165\columnwidth} >{\centering\arraybackslash}m{0.165\columnwidth} 
}
\toprule
\multirow{2}{*}{Method} & \multicolumn{2}{c}{Operational Cost} & \multicolumn{2}{c}{Overall Violation Rate} \\
\cmidrule(lr){2-3} \cmidrule(lr){4-5}
& Mean & Std. & Mean & Std.\\
\midrule
TD3        & $5.77\times10^{2}$ & $1.37\times10^{1}$ & $7.41\%$ & $\mathbf{0.160\%}$ \\
With LLM   & $\mathbf{5.55\times10^{2}}$ & $1.54\times10^{1}$ & $4.26\%$ & $0.292\%$ \\
With HITL  & $6.13\times10^{2}$ & $\mathbf{1.20\times10^{1}}$ & $\mathbf{2.53\%}$ & $0.538\%$ \\
\bottomrule
\end{tabular}
\end{table}

\begin{table}[!t]
\caption{Final test performance on 118-bus system}
\label{tab:performance_118bus}
\centering
\footnotesize
\renewcommand{\arraystretch}{1.2}
\setlength{\tabcolsep}{4pt}
\begin{tabular}{ 
>{\centering\arraybackslash}m{0.16\columnwidth} >{\centering\arraybackslash}m{0.165\columnwidth} >{\centering\arraybackslash}m{0.165\columnwidth} >{\centering\arraybackslash}m{0.165\columnwidth} >{\centering\arraybackslash}m{0.165\columnwidth} 
}
\toprule
\multirow{2}{*}{Method} & \multicolumn{2}{c}{Operational Cost} & \multicolumn{2}{c}{Overall Violation Rate} \\
\cmidrule(lr){2-3} \cmidrule(lr){4-5}
& Mean & Std. & Mean & Std.\\
\midrule
TD3        & $2.12\times10^{5}$ & $1.05\times10^{5}$ & $9.61\%$ & $1.56\%$ \\
With LLM   & $\mathbf{1.42\times10^{5}}$ & $5.70\times10^{4}$ & $9.33\%$ & $1.18\%$ \\
With HITL  & $1.78\times10^{5}$ & $\mathbf{6.45\times10^{3}}$ & $\mathbf{8.29\%}$ & $\mathbf{1.15\%}$ \\
\bottomrule
\end{tabular}
\end{table}

\begin{figure*}[!t]
\centering
\begin{minipage}[t]{0.48\textwidth}
    \centering
    \includegraphics[width=\textwidth]{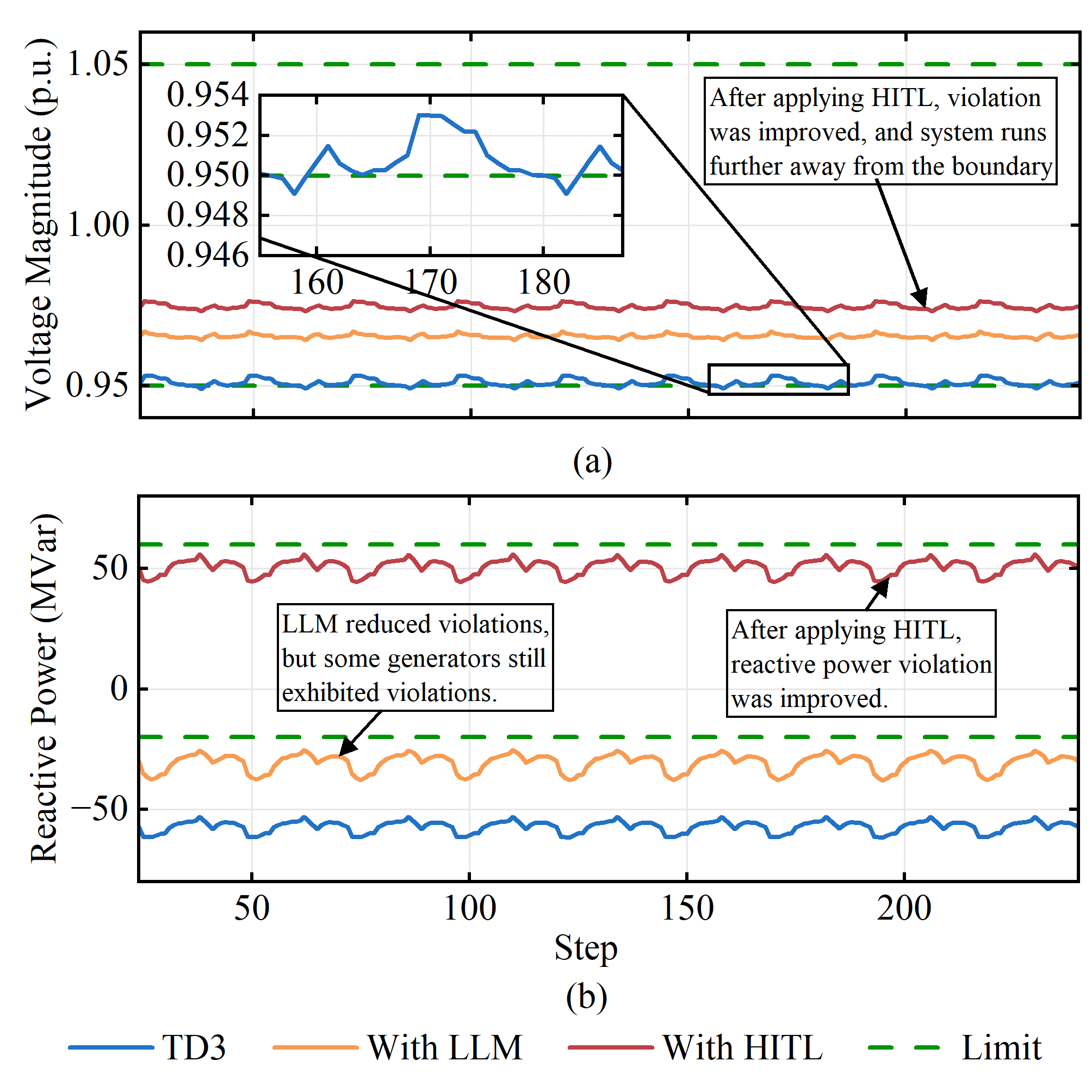}
    \caption{Operation performance under different feedback mechanisms on the 30-bus system: (a) voltage magnitude at Bus \#6; (b) reactive power output at generator \#0.}
    \label{fig:30specificvio}
\end{minipage}
\hfill
\begin{minipage}[t]{0.48\textwidth}
    \centering
    \includegraphics[width=\textwidth]{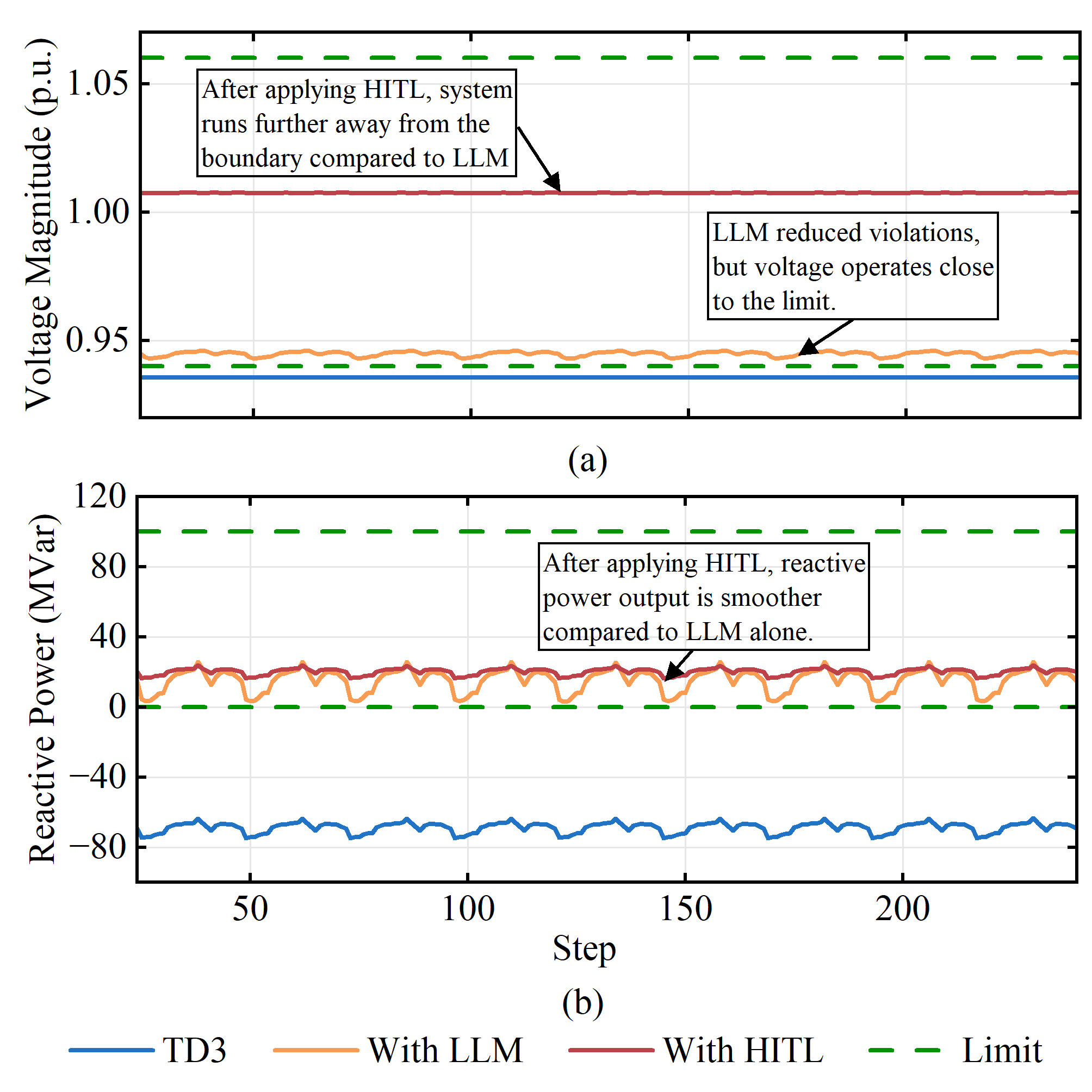}
    \caption{Operation performance under different feedback mechanisms on the 118-bus system: (a) voltage magnitude at Bus \#37; (b) reactive power output at generator \#32.}
    \label{fig:118specificvio}
\end{minipage}
\end{figure*}

\begin{figure*}[h]
\centering
\includegraphics[width=\textwidth]{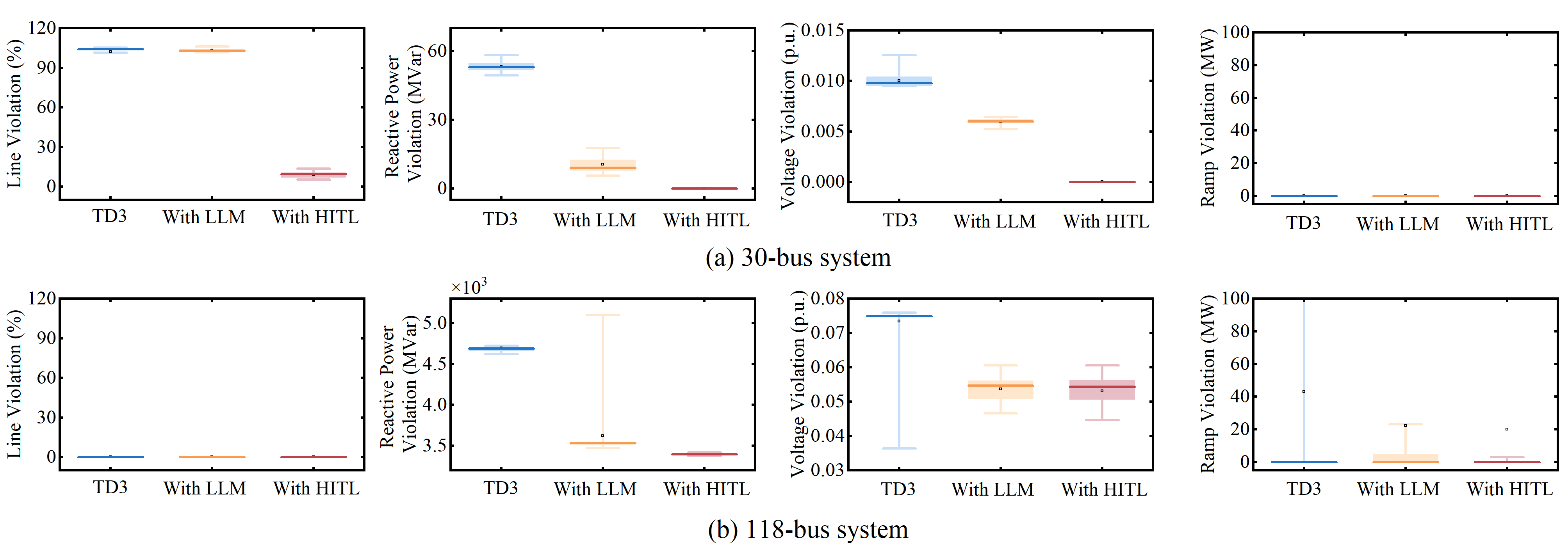}
\caption{Constraint violation comparison under different feedback mechanisms: (a) IEEE 30-bus system; (b) IEEE 118-bus system.}
\label{fig:errorbar}
\end{figure*}

The physical operation on the IEEE 30-bus system during a typical day is further illustrated in Fig.~\ref{fig:30specificvio}. As shown in Fig.~\ref{fig:30specificvio}(a), the LLM-enhanced policy already alleviates part of the voltage-related stress compared with \textit{TD3}, while the HITL-calibrated policy keeps the bus voltage farther away from the lower boundary in the representative interval. A similar trend can be observed in Fig.~\ref{fig:30specificvio}(b) for the reactive power output of Generator \#0. The \textit{with LLM} method reduces the severity of reactive power violation, but some operating points still remain close to the constraint boundary. After applying HITL calibration, the reactive power trajectory becomes more conservative and better aligned with the allowable operating region. These results indicate that the proposed calibration mechanism not only improves aggregate safety metrics, but also leads to more physically desirable operating behaviors on representative buses and generators. The effectiveness of the proposed framework on a larger transmission system is illustrated in Fig.~\ref{fig:118specificvio}. On the IEEE 118-bus system, the same qualitative trend remains valid. The \textit{with LLM} method reduces the violation level relative to TD3, while the \textit{with HITL} policy further moves the operating point away from the voltage boundary and produces a smoother reactive power response.

To further reveal how different feedback mechanisms affect specific physical constraints, Fig.~\ref{fig:errorbar} compares the constraint average violations on both systems. Relative to \textit{TD3}, the introduction of LLM feedback already reduces the mean violations in across reactive power, voltage, and ramp related constraints, indicating that language guidance can improve the agent's sensitivity to security-related operating conditions. After further incorporating HITL calibration, the violations are reduced more consistently across line, reactive power, voltage, and ramp related constraints, and the fluctuation ranges are also generally narrowed. This trend is particularly clear in the 30-bus system, where the \textit{with HITL} achieves the lowest mean violations together with smaller variability, and remains observable in the 118-bus system despite the higher dimensionality and stronger coupling of the larger network. Overall, these results show that \textit{with HITL} not only improves upon the \textit{TD3}, but also provides a clear advantage over the LLM-only method by yielding a learned RT-OPF policy that is both safer and more stable.

\section{Conclusion}
This paper proposed an LLM-enhanced deep reinforcement learning framework with human-in-the-loop for real-time optimal power flow. By introducing language evaluative feedback into offline DRL training and selectively calibrating unreliable LLM judgments through expert intervention, the proposed method establishes a reviewable and corrigible reward-shaping mechanism for RT-OPF. Case studies on the IEEE 30-bus and IEEE 118-bus systems showed that LLM feedback improves the learned policy over vanilla TD3, while the further incorporation of HITL achieves the best overall safety performance and more stable training behavior. These results demonstrate that language guidance, when combined with selective expert calibration, can improve both the safety and controllability of DRL training for power system operation without affecting the fast online deployment advantage of the learned policy.

Future work will extend the proposed framework to richer operational rules, more systematic expert trigger mechanisms, and larger-scale power system applications.

\end{document}